\documentclass[conference]{IEEEtran}
\IEEEoverridecommandlockouts

\usepackage{cite}
\usepackage{amsmath,amssymb,amsfonts}
\usepackage{algorithmic}
\usepackage{graphicx}
\usepackage{textcomp}
\usepackage{xcolor}

\def\RR{\mathbb{R}}

\def\EE{\mathbb{E}}

\def\DD{\mathcal{D}}
\def\BB{\mathcal{B}}

\def\PP{\mathbb{P}}

\usepackage{bbold}
\def\on{\mathbb{1}}

\def\piemp{\pi^{(\textsf{em})}}
\def\pitil{\pi^{(0)}}
\def\piinterp{\hat \pi^{(\textsf{ip})}}
\def\pip{\pi^{(+)}}
\def\pim{\pi^{(-)}}
\def\pig{\pi^{(g)}}

\def\gmax{g^{(max)}}

\def\thetatil{\theta^{(0)}}
\def\thetag{\theta^{(g)}}

\def\KL{d_{KL}}
\def\dTV{d_{TV}}

\def\diag{\mathsf{diag}}
\def\trace{\mathsf{trace}}

\def\pior{\pi^{(\textsf{or})}}
\def\piora{\pi^{\textsf{(or, A)}}}
\def\piorb{\pi^{\textsf{(or, B)}}}

\def\pihatl{\hat \pi^{(l)}}
\def\pihatla{\hat \pi^{(l, A)}}
\def\pihatlb{\hat \pi^{(l, B)}}

\def\Mhatla{\hat M_l^A}

\def\Bin{\textsf{Bin}}

\newtheorem{theorem}{Theorem}[section]
\newtheorem{lemma}[theorem]{Lemma}

\begin{document}

\title{Distribution Estimation with Side Information}

\author{
\IEEEauthorblockN{Haricharan Balasundaram, Andrew Thangaraj}
\IEEEauthorblockA{Department of Electrical Engineering, IIT Madras\\
                   Chennai, India\\
                   Email: haricharanb@smail.iitm.ac.in, andrew@ee.iitm.ac.in}}

\maketitle

\begin{abstract}
We consider the classical problem of discrete distribution estimation using i.i.d. samples in a novel scenario where additional side information is available on the distribution. In large alphabet datasets such as text corpora, such side information arises naturally through word semantics/similarities that can be inferred by closeness of vector word embeddings, for instance. We consider two specific models for side information--a local model where the unknown distribution is in the neighborhood of a known distribution, and a partial ordering model where the alphabet is partitioned into known higher and lower probability sets. In both models, we theoretically characterize the improvement in a suitable squared-error risk because of the available side information. Simulations over natural language and synthetic data illustrate these gains.
\end{abstract}


\section{Introduction}

Let $X^n = (X_1, X_2, \dots, X_n)$ be $n$ i.i.d. samples drawn from a probability distribution $\pi = (\pi_1, \pi_2, \dots, \pi_d)$. The problem of distribution estimation is to find an estimate $\hat \pi (X^n)$ such that $\hat \pi$ is `close' to $\pi$. In applications such as language modeling, the large alphabet scenario with $d\gg n$ makes distribution estimation inaccurate for sparsely observed or infrequent words. However, the i.i.d. sample model ignores the semantics of words that contain considerable `side' information. The main goal of this article is to develop models and study distribution estimation under an i.i.d. samples plus side information setting resulting in improved accuracy. 

Distribution estimation under the i.i.d. samples model has been studied extensively in the minimax setting with the loss function being KL-divergence \cite{kl_div_estimation_1,kl_div_estimation_2, kl_div_estimation_3}, the $\chi^2$--distance \cite{kamath15}, and the $\ell_1$--distance \cite{kamath15}. For simplicity, we focus on the squared $\ell_2$--distance studied in \cite{kamath15}, in which it is shown that the empirical frequency distribution estimate achieves an expected worst case loss of $O\left(\frac 1n \right)$ (the related add-$\sqrt{n}/d$ estimator is minimax optimal with same loss order).

In this article, we consider two models for distribution estimation with side information, study their risk/error performance theoretically and illustrate their utility in text corpora and synthetic data simulation settings. To the best of our knowledge, discrete distribution estimation appears to have not been studied in such side information models in earlier work. Some related prior work in statistical modeling of additional structure in language data include \cite{teh-2006-hierarchical,falahatgar_20}.

In the first model, we suppose that we are given a `guess' for the distribution. For instance, in the estimation of the bigram language model distribution $p(\,\cdot\,|\,\text{`Big'})$, a guess could be $p(\,\cdot\,|\,\text{`Large'})$. More formally, we assume that the true distribution is inside an $\ell_2$--ball around some known $\pitil$. We study an estimator that interpolates between the empirical distribution estimator and the `guess' $\pitil$ and derive upper bounds for the minimax risk. We derive minimax risk lower bounds by applying standard methods and compare with the upper bounds. 

In the second model, we assume knowledge of partial ordering of the distribution, i.e. we know that the symbols are partitioned into two sets $A$ (a low probability set) and $B$ (a high probability set). This is in contrast to the complete ordering assumed in \cite{birge, learning_k_modal_dist}. With this side information, we show that it is possible to utilize a two-level estimate for the species occurring $l$ times and show that it outperforms the Good-Turing one-level estimate (or natural estimate as defined in \cite{orlitsky_competitive_good_turing}) if the means of the probability masses in $A$ and $B$ are sufficiently separated.

\noindent \textbf{Organization of the Paper.} In Section~\ref{sec:notation_and_preliminaries}, we define some notation and state some preliminaries. In Section~\ref{sec:problem_formulation} we formulate the problem and state our results. We validate our results using simulations in Section~\ref{sec:simulations} and make concluding remarks in Section~\ref{sec:conclusions}. We provide proofs for the theorems in Section~\ref{sec:proofs}.

\section{Notation and Preliminaries}

\label{sec:notation_and_preliminaries}

\subsection{Notation}

Let $\DD_d$ denote the $d$--dimensional probability simplex, i.e.
\begin{equation}
    \DD_d = \{x \in \RR^{d}: \sum_{i = 1}^d x_i = 1, x_i>0\}.
\end{equation}
For any set $A \subseteq [d]$, let $\pi_A = \sum_{i \in A} \pi_i$.

Let $\on_E$ denote the indicator random variable for the event $E$. For $S \subseteq [d]$, let $N_S = N_S(X^n) \triangleq \sum_{i = 1}^n \on_{X_i \in S}$ denote the number of occurrences of elements of $S$ in the samples $X^n$. We define $N_i = N_{\{i\}}$. Next, we formally define $S_l = S_l (X^n) \triangleq \{x \in [d]: N_x(X^n) = l\}$ to be the random set of letters that have been observed exactly $l$ times in $X^n$ for $l=0,1,\ldots$. Let $\phi_l\triangleq|S_l|$. For the purpose of theoretical analysis for $l=0$, we will assume that $d$ is known so that $\phi_0$ can also be calculated exactly for every sample.

For a distribution $\pi$ and an estimator $\hat \pi$, the expected loss or risk is $\EE[L(\pi, \hat \pi (X^n))]$ for some loss function $L$. For a particular estimator, the worst-case distribution is one which maximizes the risk and the worst-case risk is
\begin{equation}
   R(\hat \pi) = \max_{\pi \in \DD_d} \EE[L(\pi, \hat \pi)].
\end{equation}
The minimax risk is the least worst-case risk achieved by any estimator, $R^* = \min_{\hat \pi} R(\hat \pi)$.

\subsection{Preliminaries}

\subsubsection{Effectiveness of the empirical estimator} We state some results regarding $R^*$ for the expected squared $\ell_2$--loss from \cite{kamath15}.

The empirical estimator $\piemp$ is defined as follows
\begin{equation}
    \piemp_i = \frac{N_i}{n}.
\end{equation}

\begin{theorem}
\label{thm:kamath_ub}
For a distribution $\pi$ and the empirical estimator $\piemp$,
\begin{equation} 
R^* \le R(\piemp) = \EE[\|\pi - \piemp (X^n)\|^2] = \frac{1 - \|\pi\|^2}{n}.
\end{equation}   
\end{theorem}
We take $\|\cdot \|$ to be the $\ell_2$ norm throughout.

While in general it is possible to take $\|\pi\| \ge 0$ in Theorem~\ref{thm:kamath_ub} to get an upper bound of $O \left(\frac 1n \right)$, we crucially use the dependence on $\|\pi\|$ in our results.

\noindent \textbf{Remark.} The empirical estimator is not optimal but rather the add--$\frac{\sqrt{n}}{d}$ estimator defined in \cite{kamath15} performs slightly better and is minimax optimal \cite[Theorem 3]{kamath15}. We utilize the empirical estimator throughout for analytical simplicity and since it is an unbiased estimator. Our results follow for the add--$\frac{\sqrt{n}}{d}$ estimator as well.

\subsubsection{Loss restricted to symbols in $S_l$} In the squared $\ell_2$--norm, note that the error decomposes as follows:
\begin{equation}
    \EE\left[\|\pi - \hat \pi\|^2 \right] = \sum_{l = 0}^n \EE \left[\sum_{i \in S_l} (\pi_i - \hat \pi_i)^2 \right]
\end{equation}
While the squared $\ell_2$--error as a whole is interesting in itself, the error restricted to symbols just in $S_l$ has also been considered in \cite{allerton_work, acharya_optimal_estimation}.

We define $Q_l(\pi, \pihatl)$ to be the error over symbols just in $S_l$ and $R_l$ to be its expected value:
\begin{equation}
    Q_l(\pi, \hat \pi) = \sum_{i \in S_l} (\pi_i - \hat \pi_i)^2, \quad R_l(\pi, \hat \pi) = \EE[Q_l(\pi, \hat \pi)].
\end{equation}
We assume that $S_l$ is non-empty.

Natural estimators assign the same probability to all letters occurring the same number of times \cite{orlitsky_competitive_good_turing}. Formally, for $x, y \in S_l$, we have $\hat \pi_x = \hat \pi_y$ for natural estimators. Almost all common estimators such as the empirical estimator \cite{canonneshortnote,han2015}, add-constant estimators studied in \cite{kamath15, mourtada2025, canonne2023} such as the Laplace estimator and the Krichevsky-Trofimov estimator \cite{kt_estimator}, and the profile-maximum likelihood estimator in \cite{orlitsky2004, acharya17} are natural.

Let $\pihatl$ be the mass assigned by a natural estimator over symbols occurring $l$ times. From this, we get
\begin{equation}
    Q_l(\pi, \pihatl) = \sum_{i \in S_l} (\pi_i - \pihatl)^2.
\end{equation}

For the `oracle' estimator which knows $\pi$ but is forced to be natural, let $\pior$ be the single probability mass it assigns to symbols occurring $l$ times. To minimize $Q_l(\pi, \pihatl)$, the minimizer is at $\pior = \frac{\sum_{i \in S_l} \pi_i}{\phi_l} = \frac{M_l}{\phi_l}$, where $M_l$ is the total mass of symbols occurring $l$ times \cite{good_turing}. Note that
\begin{IEEEeqnarray}{rCl}
    Q_l(\pi, \pihatl) &=& \sum_{i \in S_l} (\pi_i - \pior + \pior - \pihatl)^2 \nonumber\\
    & \overset{(a)}{=} & \sum_{i \in S_l} (\pi_i - \pior)^2 + \phi_l (\pior - \pihatl)^2 \label{eqn:Q_l_decomposition} \\
    & \overset{(b)}{=} & \underbrace{\sum_{i \in S_l} (\pi_i - \pior)^2}_{\text{Oracle Error}} + \underbrace{\frac{(M_l - \hat M_l)^2}{\phi_l}}_{\text{Estimator Error}}\nonumber
\end{IEEEeqnarray}
where $(a)$ follows since $\sum_{i \in S_l} (\pi^{(or)} - \pi_i) = 0$ and $(b)$ uses the definition $\hat M_l = \sum_{i \in S_l} \pihatl = \phi_l \pihatl$.

\begin{enumerate}
    \item The first term is the unavoidable oracle error characterized in \cite{allerton_work}, which provides an estimator for this unavoidable error.
    \item The second term is the estimator error. It is known that the modified Good-Turing estimator estimates the missing mass with low bias \cite{good_turing_convergence} and low minimax risk \cite{missing_mass_risk}. In particular, we note the well-known Theorem~\ref{thm:GT_MSE_iid}.
\end{enumerate}

\begin{theorem}
    For $l < n/2-1$ and $\hat M_l = \frac{l + 1}{n - l} \phi_{l + 1}$,
    \begin{equation}
         \EE[(M_l - \hat M_l)^2] \le  \frac{c_l}{n}, \label{eq:GT_MSE_iid}
    \end{equation}
    where $c_l$ is some constant that depends only on $l$.
 \label{thm:GT_MSE_iid}
\end{theorem}

\section{Problem Formulation and Results}
\label{sec:problem_formulation}

In this section, we consider two models with different side information: one with local information and the other with ordering-based information.

\subsection{Model 1}

Let $B(\pitil, \Delta)$ denote the set of valid distributions in the $\ell_2$--ball around $\pitil\in\DD_d$ of radius $\Delta \le 1$. Formally,
\begin{equation}
    B(\pitil, \Delta) = \{\pi \in \DD_d : \|\pi - \pitil\| \le \Delta\}.
\end{equation}

In this model, we suppose that $\pi \in \BB(\pitil, \Delta)$, where $\pitil$ is our `guess' distribution. Now, consider an estimator $\hat \pi (X^n, \pitil, \Delta)$ (denoted $\hat{\pi}$ in short) for $\pi$. The worst-case $\ell_2$--risk for this estimator is
\begin{equation}
    R(\pitil, \Delta, \hat \pi) = \max_{\pi \in \BB(\pitil, \Delta)} \EE[\|\pi - \hat \pi\|^2].
\end{equation}
The minimax $\ell_2$--risk for a particular $\pitil$ is 
\begin{equation}
    R^*(\pitil, \Delta) = \min_{\hat \pi} R(\pitil, \Delta, \hat \pi).
\end{equation}
We aim to come up with upper bounds and lower bounds for $R^*(\pitil, \Delta)$. 

First, we provide an upper bound for $R^*(\pitil, \Delta)$. For very small values of $\Delta$ and small values of $n$, it is better to have $\hat \pi$ closer to $\pitil$, and for large values of $\Delta$ and large values of $n$, it is better to have $\hat \pi$ closer to $\piemp$. We consider an estimator that interpolates between the empirical distribution and our guess:
\begin{equation}
    \piinterp = \alpha \piemp + (1 - \alpha) \pitil,
\end{equation}
where $\alpha \in [0, 1]$ is an estimator parameter. Such estimators are sometimes called shrinkage estimators.

\begin{theorem}
    For $\alpha = \frac{n \Delta^2}{n \Delta^2 + {1 - (\|\pitil\| - \Delta)^2}}$ in $\piinterp$, we get
    \begin{equation}
    R(\pitil, \Delta, \piinterp) \le \min \left(\Delta^2, \frac{1 - (\|\pitil\| - \Delta)^2}{n} \right).
    \end{equation}
    \label{thm:ub}
\end{theorem}
To obtain lower bounds on the minimax risk, the Dirichlet prior usually used earlier in distribution estimation problems can no longer be used directly since it might include values outside of $\BB(\pitil, \Delta)$. We provide multiple lower bounds using Le Cam's method and Assouad's lemma that are seen to be close to the upper bound in different cases.

We first provide a lower bound for the minmax risk independent of $\pitil$ and $d$ using Le Cam's method. 
\begin{theorem}
    \begin{equation}
        R^*(\pitil, \Delta) \ge \min \left(\frac{\Delta^2}{32}, \frac{\Delta}{100 n} \right) e^{-\frac 45}.
    \end{equation}
    \label{lem:lecam_lb}
\end{theorem}
Note that for $\pitil = (1, 0, \dots, 0)$, the upper bound from Theorem~\ref{thm:ub} and the lower bound from Theorem~\ref{lem:lecam_lb} are both $\Theta \left( \min \left(\Delta^2, \frac{\Delta}{n} \right) \right)$, which can be significantly better than $\Theta \left (\frac 1n \right)$ for low $\Delta$.

Next, we provide a lower bound when $\pitil$ is uniform, the proof of which utilizes Assouad's Lemma.
\begin{theorem} \label{thm:uniform_lb} For $\text{Unif}[d] = \left(\frac 1d, \frac 1d, \dots, \frac 1d \right)$ and $n \ge d$, we have
\begin{equation}
    R^*(\text{Unif}[d], \Delta) \ge \min \left(\Delta^2, \frac 1n \right) \frac{e^{-2}}{8}.
\end{equation}
\end{theorem}
Note that the upper bound from Theorem~\ref{thm:ub} and the lower bound from Theorem~\ref{thm:uniform_lb} are both $\Theta \left(\Delta^2, \frac 1n \right)$ and thus tight.

While the lower bounds and upper bounds match for the two extreme cases of $\pitil$ (deterministic and uniform), the case of general $\pitil$ remains open for improvement. Our current best lower bound uses Le Cam's method to correctly obtain the $1 - \|\pitil \|^2$ dependence as in the upper bound.
\begin{theorem}
    \begin{equation}
    R^*(\pitil, \Delta) \ge \frac{(1 - \|\pitil\|^2)}{d - 1} \min \left(\frac{\Delta^2}{12}, \frac 1{4n} \right).
    \end{equation}
    \label{thm:gen_lower_bound}
\end{theorem}
Unfortunately, the above lower bound depends on $d$ unlike the upper bound.

\subsection{Model 2}

Consider the following model for side information: we assume that the set $[d]$ is partitioned into two sets--$A$ and $B$ such that $B = [d] \setminus A$. We think of $A$ as the set of low probability symbols and $B$ as the set of high probability symbols. We define $S_l^A = S_l \cap A$, $\phi_l^A = |S_l^A|$ and similarly for $S_l^B$ and $\phi_l^B$.

For the elements in $S_l$, the new two-level estimator is allowed to assign two separate probability masses: $\pihatla$ for symbols in $S_l^A$ and $\pihatlb$ for symbols in $S_l^B$. This is \textit{not} a natural estimator but rather utilizes the side information provided. The error in this case over symbols in $S_l$ is
\begin{equation}
    Q_l (\pi, \pihatla, \pihatlb) = \sum_{i \in S_l^A} (\pi_i - \pihatla)^2 + \sum_{i \in S_l^B} (\pi_i - \pihatlb)^2.
\end{equation}

By a decomposition similar to~\eqref{eqn:Q_l_decomposition}, it can be seen that    
\begin{IEEEeqnarray}{rCl}
    \IEEEeqnarraymulticol{3}{l}{Q_l (\pi, \pihatla, \pihatlb)}
    \nonumber \\   
    &=& \sum_{i \in S_l^A} (\pi_i - \piora)^2 + \sum_{i \in S_l^B} (\pi_i - \piorb)^2 \nonumber\\
    & & + \phi_l^A (\piora - \pihatla)^2 + \phi_l^B (\piorb - \pihatlb)^2 \label{eqn:Q_l_two_level_expansion},
\end{IEEEeqnarray}
where $\piora = \frac{\sum_{S_l^A} \pi_i}{\phi_l^A}$ and $\piorb = \frac{\sum_{S_l^B} \pi_i}{\phi_l^B}$.

Note that $\pior$, $\piora$, and $\piorb$ are connected by the following algebraic identity:
\begin{equation}
    \phi_l \pior = \phi_l^A \piora + \phi_l^B \piorb. \label{eqn:oracle_decomposition}
\end{equation}

The quantity of interest to us is $Q_l (\pi, \pihatl) - Q_l (\pi, \pihatla, \pihatlb)$, which in some sense is a regret quantity measuring the improvement in performance of the two-level estimator in comparison to the one-level estimator. We note the following algebraic lemma, which crucially uses~\eqref{eqn:oracle_decomposition}.
\begin{lemma}
For any $\pihatl$, $\pihatla$, and $\pihatlb$,
\begin{IEEEeqnarray}{rCl}
    \IEEEeqnarraymulticol{3}{l}{Q_l(\pi, \pihatl) - Q_l (\pi, \pihatla, \pihatlb)} \nonumber \\   
    &=& \frac{\phi_l^A \phi_l^B}{\phi_l} (\piora - \piorb)^2 + \phi_l (\pior - \pihatl)^2 \nonumber  \\
    & & - \phi_l^A (\piora - \pihatla)^2 - \phi_l^B (\piorb - \pihatlb)^2.
\end{IEEEeqnarray}
\label{lem:error_decomposition}
\end{lemma}

We would want the $\phi_l^A (\piora - \pihatla)^2$ and $\phi_l^B (\piorb - \pihatlb)^2$ terms to be `small'. With the side information that $S_l$ is partitioned into $S_l^A$ and $S_l^B$, the two-level estimator we provide treats the problem as two separate distribution estimation problems: one on the symbols in $A$ and the other on symbols in $B$. We perform Good-Turing estimation on the elements in $A$ using $N_A$ samples and Good-Turing estimation on the elements in $B$ using $N_B$ samples. Thus, our two-level estimate is
\begin{IEEEeqnarray}{rCl}
    \pihatla &=& \frac{l + 1}{\max(1, N_A - l)} \frac{\phi_{l + 1}^A}{\phi_l^A} \\
    \pihatlb &=& \frac{l + 1}{\max(1, N_B - l)} \frac{\phi_{l + 1}^B}{\phi_l^B}. \label{eqn:two_level_estimator_defn}   
\end{IEEEeqnarray}

We note the following theorem on Good-Turing estimation of the two-level error terms.
\begin{theorem}
    For $l \le \frac n2 \min(\pi_A, \pi_B)$, we have
    \begin{multline}
        \EE \left[\phi_l^A (\piora - \pihatla)^2 + \phi_l^B (\piorb - \pihatlb)^2 \right] \\ \le \frac{c'_l}{n} \left(\frac{1}{\pi_A} + \frac{1}{\pi_B} \right),
    \end{multline}
    where $c'_l = 8l + 2c_l$ is a constant which depends only on $l$.
    \label{thm:estimation_error}    
\end{theorem}

If the probabilities in $A$ and $B$ are sufficiently apart and $n$ is sufficiently large, the two-level estimator is expected to perform well. However, if the true distribution is close to the uniform distribution, then the two-level estimate is expected to perform worse since we are using fewer samples (i.e. $N_A$ or $N_B$ instead of $n$) to estimate probabilities in each partition. This is a trade-off inherent to the problem.

Note from using $(\pior - \pihatl)^2 \ge 0$ in Lemma~\ref{lem:error_decomposition} and Theorem~\ref{thm:estimation_error} that
\begin{IEEEeqnarray}{rCl}
    \IEEEeqnarraymulticol{3}{l}{\EE[Q_l(\pi, \pihatl) - Q_l (\pi, \pihatla, \pihatlb)]} \nonumber \\   
    &\ge& \EE \left[\frac{\phi_l^A \phi_l^B}{\phi_l} (\piora - \piorb)^2 \right] - \frac{c'_l}{n} \left(\frac{1}{\pi_A} + \frac{1}{\pi_B} \right).
\end{IEEEeqnarray}

From the above, an important component of the expected gain of the two-level estimator over the one-level estimator for species in $S_l$ is seen to be
\begin{equation}
    \EE \left[\frac{\phi_l^A \phi_l^B}{\phi_l} (\piora - \piorb)^2 \right].
    \label{eqn:gain_two_level}
\end{equation}
\noindent \textbf{Remarks}: (1) While we assume initially that $A$ and $B$ are both known, it is sufficient if we know the quantities $S_l^A$, $S_l^B$, $S_{l + 1}^A$, $S_{l + 1}^B$, $N_A$, and $N_B$. (2) Lower bounds for the partial order side information is complicated by the occurrence of $\phi_l$ in \eqref{eqn:gain_two_level} and will be taken up as future work.

\section{Simulations}
\label{sec:simulations}

\begin{figure}[!ht]
    \centering
    \includegraphics[width=0.8\linewidth]{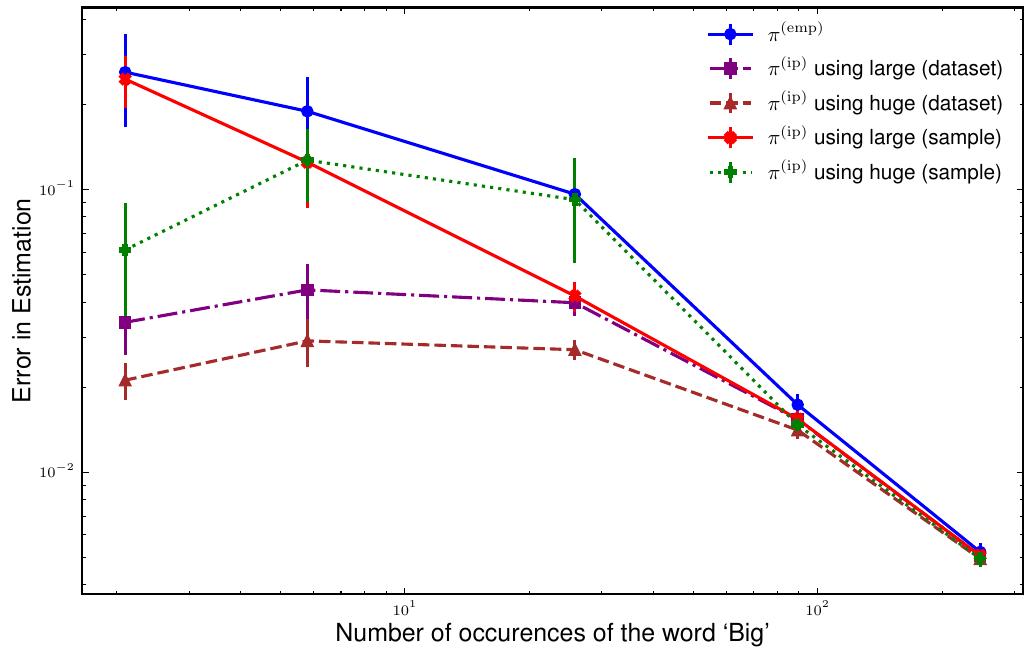}
    \caption{Estimation errors vs. number of samples for the Empirical and Interpolation Estimators for $\pitil$ from `dataset' and `sample'. All error bars are for $10$ independent repetitions.}
    \label{fig:fig_joint_plots}
\end{figure}

\noindent\textbf{Model 1}: Fig.~\ref{fig:fig_joint_plots} shows results for the estimation of the probability distribution of the words which occur after a certain word, i.e. the transition or bigram probabilities in the \texttt{IsmaelMousa/books} dataset \cite{ismaelmousa_books_dataset, datasets_library} which is a compilation of various books from Project Gutenberg \cite{gutenberg}. The target distribution $\pi$ is $p(\,\cdot\,|\,\text{`Big'})$, the transition probabilities after the word `Big' in the entire dataset. For a particular value of $n$, the samples are a contiguous section of the text containing $n$ occurrences of `Big'.

Our side information is that `big', `large', and `huge' are semantically similar. We consider various choices for $\pitil$: empirical distributions of words from the whole dataset or from the sample itself occurring after the (1) word `large', (2) word `huge'. For these cases, we take $\Delta = 1 - (\text{Cosine Similarity between \texttt{Word2Vec} \cite{word2vec} embeddings})_+$. 

From the figure, we notice that the interpolated estimators perform significantly better than the empirical estimator for small values of $n$. As $n$ grows, we observe that the interpolated estimator and the empirical estimator achieve similar losses. This is as per the theoretical predictions. We also note that using $\pitil$ from the samples, while worse than using $\pitil$ from the entire dataset, is still better than the empirical distribution estimator.

\begin{figure}[!ht]
    \centering
    \includegraphics[width=0.8\linewidth]{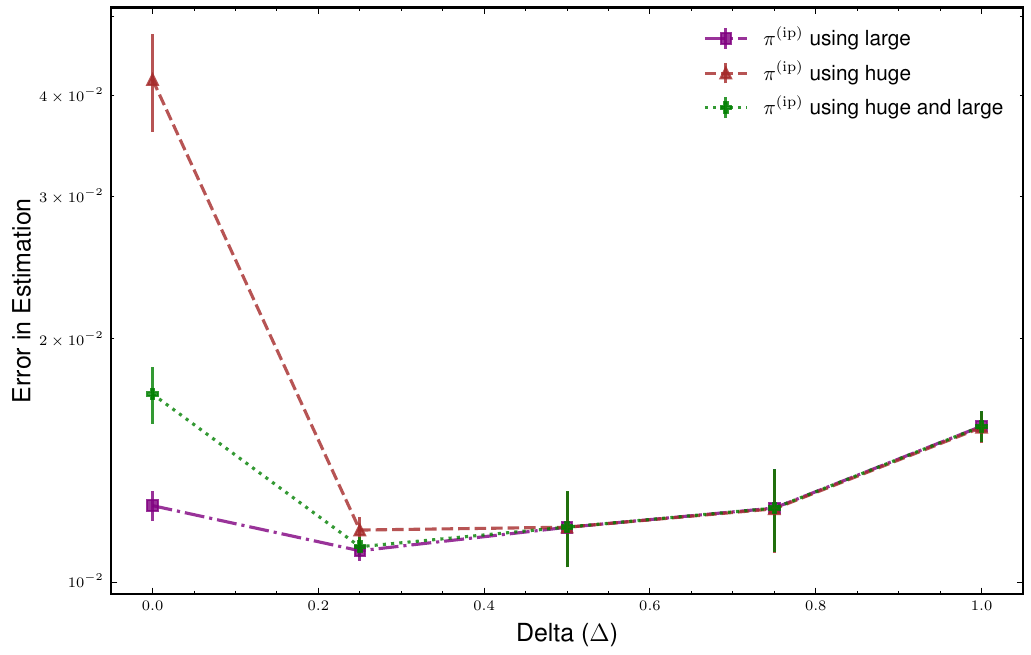}
    \caption{Estimation errors vs Delta.}
    \label{fig:fig_deltas}
\end{figure}

While we utilized the \texttt{Word2Vec} similarities to obtain the values of $\Delta$, it is unclear what choice of $\Delta$ would respect the semantics and yield optimal results. In Fig.~\ref{fig:fig_deltas}, we plot the estimation errors vs various choices of $\Delta$ for $\pitil$ being the empirical distributions of words in the whole dataset occurring after the (1) word `large', (2) word `huge', and (3) words `large' and `huge'. For the empirical estimator we take $n = 100$, i.e. $100$ appearances of the word `Big'. This shows that the parameter $\Delta$ needs to be tuned in some way using the data to improve performance, and this could possibly depend on $n$. 

\begin{figure}[!ht]
    \centering
    \includegraphics[width=0.8\linewidth]{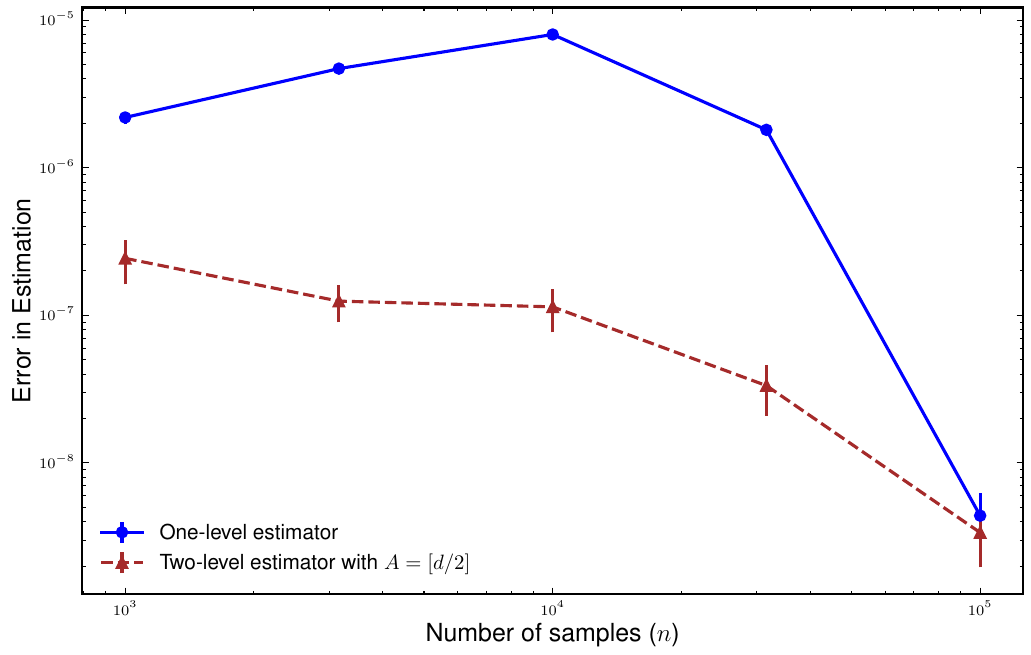}
    \caption{Two-level vs one-level estimate for the two-level distribution for $d = 10000$.}
    \label{fig:sim_two_level_distribution}
\end{figure}

\noindent\textbf{Model 2}: In Fig.~\ref{fig:sim_two_level_distribution}, we consider a synthetic distribution over $[d]$ with symbols in $[d/2]$ having probability $\frac 1{2d}$ and symbols in $[d] \setminus [d/2]$ having probability $\frac 3{2d}$. This distribution itself is two-level and the oracle estimator would achieve $0$ error. In our simulations, we observe that taking $A = [d/2]$ (note that $B = [d] \setminus A$) performs significantly better than the one-level estimator.

Next, we consider estimation of the unigram probabilities in the \texttt{IsmaelMousa/books} dataset \cite{ismaelmousa_books_dataset, datasets_library} for $l = 1$. The true unigram probabilities are computed using the entire dataset, and the sets of low and high probability are determined using a threshold value.

\begin{figure}[!ht]
    \centering
    \includegraphics[width=0.8\linewidth]{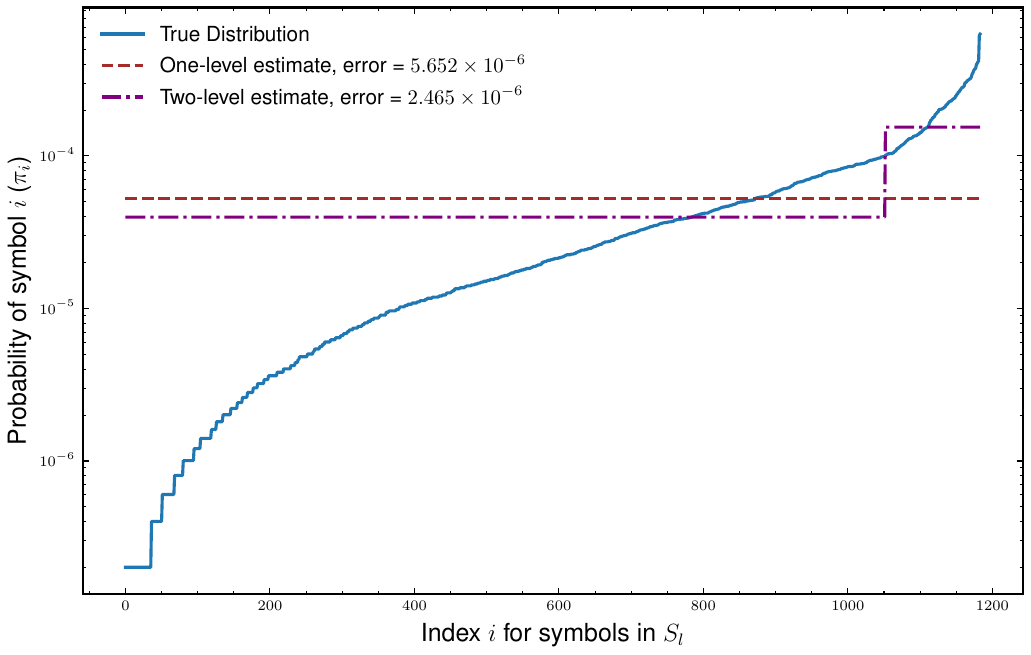}
    \caption{Distribution Estimates: Two-level vs One-level.}
    \label{fig:fig_3}
\end{figure}

In Fig.~\ref{fig:fig_3}, the true unigram probabilities for symbols in $S_1$ (contains letters that occur once) for $n = 10000$ contiguous samples are plotted and these are seen to span a wide range of values. The one- and two-level ($A = \{i: \pi_i \le 10^{-4} \}$) estimates are shown in the figure, and the error of the two-level estimate is lower by a factor of about 2.

\begin{figure}[!ht]
    \centering
    \includegraphics[width=0.8\linewidth]{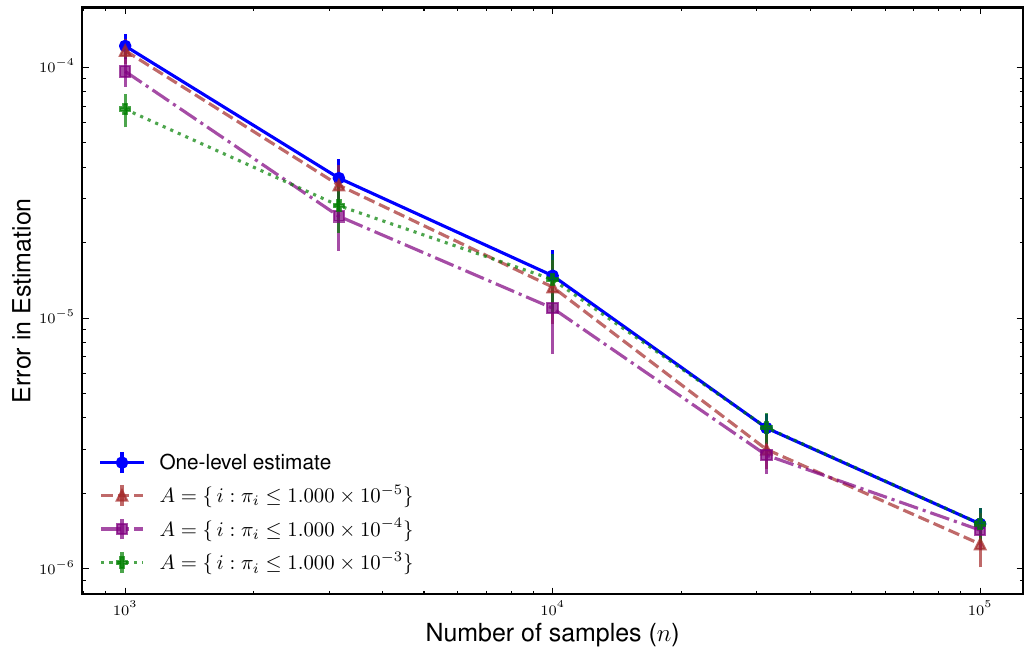}
    \caption{Error in Distribution Estimation: Two-level vs One-level.}
    \label{fig:fig_2}
\end{figure}

In Fig.~\ref{fig:fig_2}, we plot the squared error for the one- and two- estimates for the letters in $S_1$ versus $n$ for various choices of the set $A$. The two-level Good-Turing estimator performs better than the one-level estimator for appropriate values of $A$, aligning with our theoretical predictions.

\section{Conclusions}
\label{sec:conclusions}
In this work, we exploited two models of side information that could be available for distribution estimation in settings such as natural language modeling -- one based on local information using word semantics and the other on partial ordering of same-frequency symbol probabilities. We provided provable guarantees for interpolation-based estimators (for the local information case) and Good-Turing estimators (for the partial ordering case), and showed that they are in general better than those without side information. This consideration of side information in distribution estimation, perhaps for the first time, could provide an interesting new approach to statistical modeling for language and other such data with complex relationships between the words or tokens.

\section{Proofs}
\label{sec:proofs}

\subsection{Proof of Theorem~\ref{thm:ub}}
First, we show that the risk is small for even the empirical estimator since the $\max$ is over only restricted values of $\pi$.
\begin{IEEEeqnarray}{rCl}
R(\pitil, \Delta, \piemp) &=& \max_{\pi \in \BB(\pitil, \Delta)} \EE[\|\pi - \piemp\|^2] \nonumber\\
& \overset{(a)}{\le} & \max_{\pi \in \BB(\pitil, \Delta)} \frac{1 - \sum_{i=1}^d \pi_i^2}{n}
\nonumber\\
& \overset{(b)}{\le} &
\frac{1 - \left(\|\pitil\| - \Delta\right)^2}{n} \label{eqn:emp_risk_ub},
\end{IEEEeqnarray}
where $(a)$ comes from \cite[Section 2]{kamath15} and $(b)$ follows since $\|\pi\|^2 \ge (\|\pitil\| - \Delta)^2 $ from the reverse triangle inequality for $\pi \in \BB(\pitil, \Delta)$.

\begin{IEEEeqnarray}{rCl}
\IEEEeqnarraymulticol{3}{l}{R(\pitil, \Delta, \piinterp)} \nonumber \\
& = & \max_{\pi \in \BB(\pitil, \Delta)} \EE[\|\pi - \piinterp (X^n)\|^2] \nonumber\\
& \overset{(a)}{=} & \max_{\pi \in \BB(\pitil, \Delta)} \EE[\|\alpha (\pi - \piemp) + (1 - \alpha) (\pi - \pitil) \|^2] \nonumber\\
& \overset{(b)}{=} & \max_{\substack{\pi \in \\\BB(\pitil, \Delta)}} \left(\alpha^2 \EE[\|\pi - \piemp\|^2] + (1 - \alpha)^2 \EE[\|\pi - \pitil\|^2] \right) \nonumber \\
& \overset{(c)}{\le} & \alpha^2 \frac{1 - (\|\pitil\| - \Delta)^2}{n} + (1 - \alpha)^2 \Delta^2, \label{eqn:ub_to_be_minimized}
\end{IEEEeqnarray}
where $(a)$ follows from by expanding $\piinterp$, $(b)$ follows since $\EE[(\pi - \piemp)^T (\pi - \pitil)] = 0$ since the empirical estimator is an unbiased estimator, and $(c)$ follows from \eqref{eqn:emp_risk_ub}.

Next, notice that the minimizer of \eqref{eqn:ub_to_be_minimized} occurs at
\begin{equation}
    \alpha = \frac{\Delta^2}{\Delta^2 + \frac{1 - (\|\pitil\| - \Delta)^2}{n}}
\end{equation}
and the minimum value is:
\begin{equation}
    R(\pitil, \Delta, \piinterp) \le \frac{\Delta^2 (1 - (\|\pitil\| - \Delta)^2)/n}{\Delta^2 + (1 - (\|\pitil\| - \Delta)^2)/n}.
\end{equation}
Using $\frac{AB}{A + B} \le \min(A, B)$ gives the desired expression.

\subsection{Proof of Theorem~\ref{lem:lecam_lb}}

We state the lemma for Le Cam's two-point method \cite{bin_yu}.
\begin{lemma}
    \label{lem:lecam}
    \begin{multline*}
            R^*(\pitil, \Delta) \ge \\  \max_{\substack{\pip, \pim \\ \in \BB(\pitil, \Delta)}} \frac{(\pip - \pim)^2 (1 - \dTV(\pip, \pim))}{4}.        
    \end{multline*}
\end{lemma}
The distributions $\pip$ and $\pim$ are usually constructed by perturbing some of the large-valued coordinates of a base distribution in $\BB(\pitil, \Delta)$. To ensure the existence of such distributions, we provide Lemma~\ref{lem:mass_rearrangement}.
\begin{lemma}
    \label{lem:mass_rearrangement}
    For every $\pitil$, $\exists {\pitil}'$ such that
    \begin{enumerate}
        \item $\|{\pitil}' - \pitil\|^2 \le \frac{\Delta^2}{2}$.
        \item There are two coordinates $i$ and $j$ in ${\pitil}'$ that are both greater than $\frac{\Delta}{\sqrt{12}}$.
    \end{enumerate}
\end{lemma}

\begin{IEEEproof} We pick an $i$ and $j$ arbitrarily and we `move' a probability mass from other indices to these indices. It is always possible to move probability masses from other indices since the masses sum up to $1$. Let $a_i$ and $a_j$ be the probability masses required to be added to indices $i$ and $j$ respectively to make them $\ge \frac{\Delta}{\sqrt{12}}$, i.e. 
\begin{equation}
a_i = \left(\frac{\Delta}{\sqrt{12}} - \pi_i \right)_+,
\qquad
a_j = \left(\frac{\Delta}{\sqrt{12}} - \pi_j \right)_+ . \end{equation}
Here, $(x)_+ = \max(x, 0)$.

Clearly $0 \le a_i, a_j \le \frac{\Delta}{\sqrt{12}}$. From the remaining indices, we move a total probability mass of exactly $a_i + a_j$. Let the new distribution after this movement be ${\pitil}'$.

Next, consider $\|{\pitil}' - \pitil\|^2$. Observe that 
\begin{IEEEeqnarray}{rCl}
\|{\pitil}' - \pitil\|^2
&\overset{(a)}{\le}& a_i^2 + a_j^2 + (a_i + a_j)^2 \nonumber\\
&=& 2 \cdot \left(a_i^2 + a_j^2 + a_i a_j \right) \le \frac{\Delta^2}{2},\nonumber
\end{IEEEeqnarray}
where $(a)$ follows since ${\pitil}'_i \ge \pitil_i$ and ${\pitil}'_j \ge \pitil_j$ and for the other indices $k \neq i, k \neq j$, ${\pitil}'_k \le \pitil_k$. \end{IEEEproof}

Using Lemma~\ref{lem:mass_rearrangement}, we obtain a ${\pitil}'$ as defined above. Now consider two distributions
\begin{equation}
    \pi^{(\pm)} = {\pitil}' \pm \tau(e_i - e_j).
\end{equation}
Here $e_k = (0,\ldots,0,1,0,\ldots,0)$, where the $1$ appears in the $k$-th coordinate.

\begin{enumerate}
    \item To ensure non-negativity of coordinates, we need $\tau \le \frac{\Delta}{\sqrt{12}}$.
    \item To ensure $\pi^{(\pm)} \in B(\pitil, \Delta)$, we need $\|\pi^{(\pm)} - \pitil\| \le \|\pi^{(\pm)} - {\pitil}'\| + \|\pitil - {\pitil}'\| \le \sqrt{2} \tau + \frac{\Delta}{\sqrt{2}}$ to be at most $\Delta$. We take $\tau \le \frac{\Delta}{\sqrt{32}}$.
    \item Observe that $\|\pip - \pim\|^2 = 8\tau ^2$.
    \item Also observe that 
        \begin{equation}
            \KL(\pip \| \pim) \le 4 \tau^2 \left(\frac{1}{\pi_i - \tau} + \frac{1}{\pi_j - \tau} \right).
        \end{equation}
    Note that $\pi_i - \tau \ge \frac{\Delta}{\sqrt{12}} - \frac{\Delta}{\sqrt{32}} \ge \frac{\Delta}{10}$, and likewise for $\pi_j - \tau$. Using this, note that  
        \begin{equation}
            \KL(\pip \| \pim) \le \tau^2 \frac{80}{\Delta}.
        \end{equation}
\end{enumerate}

Next, we use Bretagnolle-Huber bound \cite[Theorem 1]{cannone_kl_tv} in Lemma~\ref{lem:lecam} to get a lower bound for $0 \le \tau \le \frac{\Delta}{\sqrt{32}}$:
\begin{IEEEeqnarray}{rCl}
    R^*(\pitil, \Delta) &\ge& \max_{\substack{\pip, \pim \\ \in \BB(\pitil, \Delta)}} \frac{(\pip - \pim)^2}{8} e^{-n \KL(\pip, \pim))} \nonumber \\
    &\ge& \max_{\tau \le \frac{\Delta}{\sqrt{32}}}\tau^2 e^{-n \tau^2 \frac{80}{\Delta}}.
\end{IEEEeqnarray}
Taking $\tau = \min \left(\frac{\Delta}{\sqrt{32}}, \frac {1}{10} \sqrt{\frac {\Delta}{n} } \right)$ and using $n \tau^2 \frac{80}{\Delta} \le \frac 45$, we get the desired result.

\subsection{Proof of Theorem~\ref{thm:uniform_lb}}

First, we state Assouad's Lemma \cite{bin_yu}.
\begin{lemma}
\label{lem:assouad}
Consider $\theta_v$'s, where $v \in \{-1, 1\}^l$ and
\begin{equation}
    \label{eqn:condition_assouad}
    \min_{a \in A} L(\theta_v, a) + L(\theta_{v'}, a) \ge \lambda \cdot d_H(v, v').
\end{equation}   

Then,
\begin{equation}
    R(\hat \pi) \ge \frac{l \lambda}{4} e^{ - n \cdot \max_{v, v': d_H(v, v') = 1} \KL(\theta_v, \theta_{v'})}.
\end{equation}
\end{lemma}

Without loss of generality, we assume that $d$ is even. In utilizing this lemma, we take $l = \frac d2$, $\mathcal{A} = \BB(\pitil, \Delta)$, and the loss $L$ to be the expected squared $\ell_2$--norm. Clearly, $v \in \{-1, +1\}^{\frac d2}$. Define $\theta_v$ as follows:
\begin{equation}
    \theta_v = \left(\frac 1d + \tau v_1, \frac 1d - \tau v_1, \dots, \frac 1d + \tau v_{\frac d2}, \frac 1d - \tau v_{\frac d2} \right).
\end{equation}

\begin{enumerate}
    \item To ensure non-negativity of coordinates, we need $\tau \le \frac 1d$. It can be seen that $\theta_v$ is a valid distribution.
    \item To ensure that $\theta_v \in \BB(\thetatil, \Delta)$, note that we need $\|\theta_v - \thetatil\| = \tau \sqrt{d} \le \Delta$, i.e. $\tau \le \frac{\Delta}{\sqrt{d}}$.
    \item To ensure that \eqref{eqn:condition_assouad} is satisfied in Lemma~\ref{lem:assouad}, note that
    \begin{IEEEeqnarray}{rCl}
        \min_{a \in \mathcal{A}} \| \theta_v - a \|^2 + \| a - \theta_v' \|^2 
         & \ge & \frac{\| \theta_v - \theta_v' \|^2 }{2} \nonumber\\
         & = & \frac{d_H (v, v')}{2} (4 \tau^2 + 4 \tau^2) \nonumber \\
         & = &d_H (v, v') 4 \tau^2\nonumber     
    \end{IEEEeqnarray}
    It can be seen that \eqref{eqn:condition_assouad} holds for $\lambda = 4 \tau^2$.
    \item Next, we compute 
        \begin{equation}\max_{v, v': d_H(v, v') = 1} \KL(\theta_v, \theta_{v'}).\end{equation}
    Since the Hamming distance is at most $1$, $v$ and $v'$ differ only at one coordinate, say $i$. Thus, $\theta_v$ and $\theta_{v'}$ would be different only at coordinates $2 i - 1$ and $2 i$. Take $\tau d \le \frac 12$. Note that
\begin{IEEEeqnarray}{rCl}
    \IEEEeqnarraymulticol{3}{l}{\KL(\theta_v, \theta_{v'})} \nonumber\\
    &\overset{(a)}{=}& \left(\frac 1d + \tau \right) \log \left(\frac{\frac 1d + \tau}{\frac 1d - \tau} \right) + \left(\frac 1d - \tau \right) \log \left(\frac{\frac 1d - \tau}{\frac 1d + \tau} \right) \nonumber \\
    &=& 2 \tau \log \left(\frac{\frac 1d + \tau}{\frac 1d - \tau} \right) \nonumber\\
    & \overset{(b)}{\le } & 2 \tau \cdot 4 d \tau,\nonumber
\end{IEEEeqnarray}
where $(a)$ follows since $\theta_v$ and $\theta_v$ are different only at two coordinates and $(b)$ follows since $\log(\frac{1 + x}{1 - x}) \le 4 x$ for $x \le \frac 12$. Thus, it can be seen that the KL--divergence is at most $8 d \tau^2$.
\end{enumerate}

Using Assouad's Lemma, we see that

\begin{equation}
    R^*(\pitil, \Delta) \ge \frac d2 \frac{4 \tau^2}{4} e^{-8 nd \tau^2} = \frac d2 \tau^2 e^{-8 nd \tau^2}. 
\end{equation}
We take $\tau^2 = \frac{\min(\Delta^2, \frac 1n)}{4 d}$. Clearly, $\tau \le \frac{\Delta}{\sqrt{d}}$ and $8 n d \tau^2 \le 2$. To ensure $\tau \le \frac 1{2d}$, we take $n \ge d$. From this, we get the desired theorem.

\subsection{Proof of Theorem~\ref{thm:gen_lower_bound}}
The proof uses a Le Cam style two-point method in square-root coordinates and uses Hellinger distance in the lower bound. Define $\thetatil := \sqrt{\pitil}$ \footnote{Here, $\sqrt{}$ refers to the element-wise square root.} and let $W := \diag(\pi)$ be the $d\times d$ matrix with $\pi$ on the diagonal. Let $g$ be a unit vector such that $(\thetatil)^T g = 0$. Let $\thetag = \thetatil + \tau g$, where $\tau \le 1$.

\begin{lemma}
    \begin{equation}
        \|\thetag\|^2 = 1 + \tau^2.
    \end{equation}
\end{lemma}

\begin{IEEEproof}
    \begin{IEEEeqnarray}{rCl}
        \|\thetag\|^2 &=& \sum_{i = 1}^d (\thetag_i)^2 \nonumber\\ 
        &=& \sum_{i = 1}^d ((\thetatil_i)^2 + 2 \tau g_{i} \thetatil_i + \tau^2 g_i^2) \nonumber\\ 
        &\overset{(a)}{=}& 1 + \tau^2,    
    \end{IEEEeqnarray}
where $(a)$ follows since $\sum_{i = 1}^d g_i \thetatil_i = 0$ and $\sum_{i = 1}^d g_i^2 = 1$. \end{IEEEproof}

Let $\pig = \frac{(\thetag)^2}{1 + \tau^2}$. Non-negativity and summation to $1$ follow from the definition, and thus $\pig \in \DD_d$.

To ensure $\pig \in \BB(\pitil, \Delta)$, observe that we have:
\begin{IEEEeqnarray}{rCl}
    \pig_i - \pitil_i &=& \frac{2 \tau g_i \thetatil_i + \tau^2 \left(g_i^2 - (\thetatil_i)^2 \right)}{1 + \tau^2} \label{eqn:diff_expansion} \\
    (\pig_i - \pitil_i)^2 &\overset{(a)}{\le}& \frac{8 \tau^2 g_i^2 (\thetatil_i)^2 + 2 \tau^4 \left(g_i^2 - (\thetatil_i)^2 \right)^2}{(1 + \tau^2)^2} \nonumber \\
    \sum_{i = 1}^d (\pig_i - \pitil_i)^2 &\overset{(b)}{\le}& \sum_{i = 1}^d \frac{8 \tau^2 g_i^2 + 2 \tau^4 \left(g_i^4 + (\thetatil_i)^4 \right)}{(1 + \tau^2)^2} \nonumber\\
    &\overset{(c)}{\le}& \frac{8 \tau^2 + 4 \tau^4}{(1 + \tau^2)^2} \nonumber\\
    &\overset{(d)}{\le}& 12 \tau^2 \le \Delta^2 \text{ for } \tau \le \frac{\Delta}{\sqrt{12}}, \label{eqn:squared_l2_norm_proof}
\end{IEEEeqnarray}
where $(a)$ follows since $(A+B)^2 
\le 2A^2 + 2B^2$, $(b)$ follows from dropping the negative terms and using $(\thetatil_i)^2 \le 1$, $(c)$ follows from Cauchy-Schwarz, and $(d)$ follows since $\tau^4 \le \tau^2$ and $(1 + \tau^2)^2 \ge 1$.

Let $g_1, g_2, \dots, g_{d - 1}$ be the basis for the vector space $\{u \in \RR^d: u^T \thetatil = 0\}$.

\begin{lemma}\label{lem:eigenvalue}
\begin{equation}
\sum_{k=1}^{d-1} g_k^T W g_k = 1 - \|\pitil \|^2.    
\end{equation}
\end{lemma}

\begin{IEEEproof} Let $G_{d \times (d - 1)} = [g_1| \cdots|g_{d-1}]$. Note that $GG^T = I - \thetatil (\thetatil)^T$ since they are both orthogonal projection matrices.
\begin{IEEEeqnarray}{rCl}
\sum_{k = 1}^{d - 1} g_k^T W g_k & = & \trace (W GG^T) \nonumber\\
   & = &\trace(W) - \trace(W \thetatil (\thetatil)^T) \nonumber\\
   & = & 1 - \sum_{k = 1}^{d - 1} (\pitil_i)^2. \nonumber   
\end{IEEEeqnarray}

\end{IEEEproof}

From Lemma~\ref{lem:eigenvalue}, note that $\exists$ a $\gmax$ such that
\begin{equation}
    (\gmax)^T W \gmax \ge \frac{1-\|\pitil\|^2}{d-1}.
    \label{eqn:max_g}
\end{equation}

\noindent \textbf{Distributions in Le Cam lower bound:} Let $\pip = \pi^{(\gmax)}$ and $\pim = \pi^{(-\gmax)}$. Note that $\pip, \pim \in \BB(\pitil, \Delta)$.
\begin{enumerate}
    \item $\ell_2$--norm separation:
\begin{IEEEeqnarray}{rCl}
    \|\pip - \pim\|^2 &\overset{(a)}{=}& \sum_{k = 1}^{d - 1} \left(\frac{4 \tau \gmax_i \thetatil_i}{1 + \tau^2} \right)^2 \nonumber\\
    &=& \sum_{k = 1}^{d - 1} \left(\frac{16 \tau^2 (\gmax_i)^2 \pitil_i}{\left(1 +\tau^2\right)^2} \right)\nonumber\\
    &\overset{(b)}{=}& 16 \frac{\tau^2}{(1 + \tau^2)^2} (\gmax)^T W \gmax \nonumber\\
    &\ge& 16 \tau^2 \frac{(1 - \|\pitil\|^2)}{d - 1},\nonumber
\end{IEEEeqnarray}
where $(a)$ follows from \eqref{eqn:diff_expansion}, $(b)$ follows from a rearrangement of the equation in matrix form, and $(c)$ follows from \eqref{eqn:max_g}.

\item Bound on the $\KL$: We first bound the Hellinger distance. Note that $H(\pip, \pim) = \|\sqrt{\pip} - \sqrt{\pim}\| \le \frac{2 \tau \|\gmax\|}{\sqrt{1 + \tau^2}} \le 2 \tau$. Next, note from \cite[Section 2.4]{tsybakov} that 
\begin{IEEEeqnarray}{rCl}
    \IEEEeqnarraymulticol{3}{l}{H^2((\pip)^{\otimes n}, (\pim)^{\otimes n})}\nonumber\\ 
    &=& 2 \left ( 1 - \left(1 - \frac{H^2(\pip, \pim)}{2} \right)^n \right) \nonumber \\
    &\overset{(a)}{\le} & 2 n H^2(\pip, \pim) \le 2 n \tau^2,\nonumber
\end{IEEEeqnarray}
where $(a)$ follows from $1 - (1 - x)^n \le nx$. Finally, we utilize the bound $d_{TV}(P, Q) \le \sqrt{\frac{H(P, Q)}{2}}$ to get $\KL(\pip, \pim) \le \sqrt{n} \tau$.
\end{enumerate}

We next use Theorem~\ref{lem:lecam} to get
\begin{equation}
    R^*(\pitil, \Delta) \ge \max_{0 \le \tau \le \frac{\Delta}{\sqrt{12}}} 2 \tau^2 \frac{(1 - \|\pitil\|^2)}{d - 1} (1 - \sqrt{n} \tau).
\end{equation}
We pick $\tau = \min \left(\frac{\Delta}{\sqrt{12}}, \frac{1}{2 \sqrt{n}} \right)$ to get the desired result.

\clearpage

\subsection{Proof of Lemma~\ref{lem:error_decomposition}}

First, note from~\eqref{eqn:Q_l_decomposition} and~\eqref{eqn:Q_l_two_level_expansion} that
\begin{IEEEeqnarray}{rCl}
    \IEEEeqnarraymulticol{3}{l}{Q_l(\pi, \pihatl) - Q_l (\pi, \pihatla, \pihatlb)}
    \nonumber \\   
    &=&  \sum_{i \in S_l} (\pi_i - \pior)^2 + \phi_l (\pior - \pihatl)^2  \nonumber \\
    & & - \sum_{i \in S_l^A} (\pi_i - \piora)^2 - \sum_{i \in S_l^B} (\pi_i - \piorb)^2 \nonumber\\
    & & - \phi_l^A (\piora - \pihatla)^2 - \phi_l^B (\piorb - \pihatlb)^2 \label{eqn:error_decomposition_0}.
\end{IEEEeqnarray}

We will first bound the oracle error terms. Note from~\eqref{eqn:Q_l_decomposition} that
\begin{equation*}
    \sum_{i \in S_l^A} \left((\pi_i - \pior)^2 - (\pi_i - \piora)^2 \right) =  \phi_l^A (\piora - \pior)^2
\end{equation*}
and that a similar decomposition follows for the elements in $S_l^B$ also. From this, we get
\begin{IEEEeqnarray}{rCl}
    \IEEEeqnarraymulticol{3}{l}{\sum_{i \in S_l} (\pi_i - \pior)^2 - \sum_{i \in S_l^A} (\pi_i - \piora)^2 - \sum_{i \in S_l^B} (\pi_i - \piorb)^2} \nonumber \\
    &= & \phi_l^A (\piora - \pior)^2 + \phi_l^B (\piorb - \pior)^2.
    \label{eqn:error_decomposition_1}
\end{IEEEeqnarray}
Next, consider $\phi_l^A (\piora - \pior)^2$. Note from~\eqref{eqn:oracle_decomposition} and $\phi_l = \phi_l^A + \phi_l^B$ that $\piora - \pior = \piora - \frac{\phi_l^A \piora + \phi_l^B \piorb}{\phi_l} = \frac{\phi_l^B}{\phi_l} (\piora - \piorb)$. From this, we get
\begin{IEEEeqnarray}{rCl}
    \phi_l^A (\piora - \pior)^2 & = & \frac{(\phi_l^B)^2 \phi_l^A}{\phi_l^2} (\piora - \piorb)^2.\nonumber
\end{IEEEeqnarray}
Using this expression in~\eqref{eqn:error_decomposition_1} we get
\begin{IEEEeqnarray}{rCl}
    \IEEEeqnarraymulticol{3}{l}{\sum_{i \in S_l} (\pi_i - \pior)^2 - \sum_{i \in S_l^A} (\pi_i - \piora)^2 - \sum_{i \in S_l^B} (\pi_i - \piorb)^2} \nonumber \\
    &= & \frac{(\phi_l^B)^2 \phi_l^A}{\phi_l^2} (\piora - \piorb)^2 + \frac{(\phi_l^A)^2 \phi_l^B}{\phi_l^2} (\piora - \piorb)^2 \nonumber \\
    &= & \frac{\phi_l^A \phi_l^B}{\phi_l} (\piora - \piorb)^2
\end{IEEEeqnarray}

Substituting this in \eqref{eqn:error_decomposition_0} gives the lemma directly.

\subsection{Proof of Theorem~\ref{thm:estimation_error}}

First, observe that
\begin{multline*}
    \EE[\phi_l^A (\piora - \pihatla)^2] \\ \le \EE[(\phi_l^A)^2 (\piora - \pihatla)^2] \le \EE[(M_l^A - \Mhatla)^2],
\end{multline*}
where $\Mhatla = \frac{l + 1}{\max(1, N_A - l)} \phi_{l + 1}^A = \phi_l^A \pihatla$.

Out of the $n$ samples, note that $N_A$ and $N_B$ are the number of samples from the sets $A$ and $B$ respectively. Clearly, $N_A + N_B = n$ and $N_A \sim \Bin(n, \pi_A)$ and $N_B \sim \Bin(n, \pi_B)$. The remainder of the proof bounds $\EE[(M_l^A - \Mhatla)^2]$.

First, note from Theorem~\ref{thm:GT_MSE_iid} and $(l + 1)\phi_{l + 1}^A \le N_A$ that 
\begin{equation*}
    \EE[(M_l^A - \Mhatla)^2 | N_A = n_A] \le \begin{cases}
    \displaystyle \frac{c_l}{n_A}, & n_A \ge l + 1 \\
    n_A, &\text{otherwise}. \end{cases}
\end{equation*}
A slightly looser bound yields
\begin{equation}
    \EE[(M_l^A - \Mhatla)^2 | N_A = n_A] \le \begin{cases}
    \displaystyle \frac{2 c_l}{n_A + 1}, & n_A \ge l + 1 \\
    n_A, &\text{otherwise}. \end{cases} \label{eqn:cases_N_A}
\end{equation}

Now, note that
\begin{IEEEeqnarray}{rCl}
    \IEEEeqnarraymulticol{3}{l}{\EE[(M_l^A - \Mhatla)^2]} \nonumber\\
    & \overset{(a)}{=} & \sum_{n_A = 0}^n \EE[(M_l^A - \Mhatla)^2 | N_A = n_A] \, \PP(N_A = n_A) \nonumber\\
    &\overset{(b)}{\le} & l \, \PP(N_A \le l) + \EE \left[\frac{2 c_l}{N_A + 1} \right] \nonumber\\
    &\overset{(c)}{\le} & l \, e^{- \frac{n \pi_A}{8}} + \EE \left[\frac{2 c_l}{N_A + 1} \right] \nonumber\\
    &\overset{(d)}{\le} & \frac{8 l}{n \pi_A} + 2 c_l \frac{(1 - (1 - \pi_A)^{n + 1})}{(n + 1) \pi_A}  \nonumber\\
    &\overset{(e)}{\le} & \frac{8 l}{n \pi_A} + 2 c_l \frac{1}{n \pi_A} \overset{(f)}{\le} \frac{c'_l}{n \pi_A}\nonumber
\end{IEEEeqnarray}
where $(a)$ follows from the theorem of total probability, $(b)$ uses~\eqref{eqn:cases_N_A}, $(c)$ follows from $l \le \frac{n \pi_A}{2}$ in the Chernoff bound, $(d)$ uses $e^{-x} \le \frac 1x$ and standard expected value calculations, $(e)$ uses $(1 - \pi_A)^{n + 1} \ge 0$ and $\frac{1}{n + 1} \le \frac{1}{n}$, and $(f)$ defines $c'_l = 8l + 2c_l$ to be some constant.

Applying a similar procedure for the set $B$ gives the desired theorem.

\clearpage

\bibliographystyle{IEEEtran}
\bibliography{refs}

\end{document}